\documentclass[10pt,final,doublecolumn]{IEEEtran}
\usepackage{amsmath}
\usepackage{latexsym}
\usepackage{graphicx}
\usepackage{bbding}
\usepackage{enumerate}
\usepackage{multicol}
\usepackage{color}
\usepackage{slashbox}
\IEEEoverridecommandlockouts
\begin{document}

\title{On Interference Alignment with Imperfect CSI: Characterizations of Outage Probability, Ergodic Rate and SER}
\author{\normalsize
Xiaoming~Chen, \emph{Senior Member, IEEE}, and Chau~Yuen,
\emph{Senior Member, IEEE}
%\thanks{This work was supported by the Natural
%Science Foundation of China (No. 61301102), the Natural Science
%Foundation of Jiangsu Province (No. BK20130820), the Doctoral Fund
%of Ministry of Education of China (No. 20123218120022), and also
%Singapore University of Technology and Design.}
\thanks{Xiaoming~Chen ({\tt chenxiaoming@nuaa.edu.cn}) is with the
College of Electronic and Information Engineering, Nanjing
University of Aeronautics and Astronautics, Nanjing 210016, China.
Chau~Yuen ({\tt yuenchau@sutd.edu.sg}) is with the Singapore
University of Technology and Design, Singapore.}} \maketitle

\begin{abstract}
In this paper, we give a unified performance analysis of
interference alignment (IA) over MIMO interference channels. Rather
than the asymptotic characterization, i.e. degree of freedom (DOF)
at high signal-to-noise ratio (SNR), we focus on the other practical
performance metrics, namely outage probability, ergodic rate and
symbol error rate (SER). In particular, we consider imperfect IA due
to the fact that the transmitters usually have only imperfect
channel state information (CSI) in practical scenario. By
characterizing the impact of imperfect CSI, we derive the exact
closed-form expressions of outage probability, ergodic rate and SER
in terms of CSI accuracy, transmit SNR, channel condition, number of
antennas, and the number of data streams of each communication pair.
Furthermore, we obtain some important guidelines for performance
optimization of IA under imperfect CSI by minimizing the performance
loss over IA with perfect CSI. Finally, our theoretical claims are
validated by simulation results.
\end{abstract}


\section{Introduction}
Interference is always the bottleneck for performance improvement in
an interference networks, e.g. multi-cell network \cite{Multicell1}
\cite{Multicell2} and ad-hoc network \cite{AdHoc1} \cite{AdHoc2}.
Various advanced interference mitigation techniques, such as
coordinated beamforming over MIMO interference channel
\cite{CoordinatedBeamforming1}-\cite{CoordinatedBeamforming3} (and
the references therein), have been proposed to optimize the system
performance. In particular, interference alignment (IA) receives
considerable attentions, as it is able to exploit the maximum
degrees of freedom (DOF) by aligning all the interferences into a
specific area, so as to support higher number of concurrent data
streams that will be free from interference \cite{IA1}-\cite{IA3}.

\subsection{Related Works}
Performance analysis is an important precondition for the design of
system parameters, and hence provides useful guidelines for
performance optimization \cite{PerformanceAnalysis1}
\cite{PerformanceAnalysis2}. Since the channel capacity of
interference channel is still unknown, performance analysis for IA
becomes challenging. In this context, most works analyze the
multiplexing gain at high signal-to-noise ratio (SNR), namely degree
of freedom (DOF) \cite{DOF1}-\cite{DOF3}. It is proved that IA can
achieve at most $KM/2$ DOFs over MIMO interference channels with $K$
transmitter-receiver pairs each employing $M$ antennas \cite{IA1}.
Then, the generalized DOF region is obtained for the general case of
the MIMO Gaussian interference channel with arbitrary number of
antennas at each node and the SNRs vary with arbitrary exponents to
a nominal SNR \cite{GDOF}. However, DOF is an asymptotic performance
metric, which is not able to make accurate predictions about the
behavior of channel capacity at low and moderate SNRs under
practical range of operation. Aiming to solve this problem, the
authors in \cite{CapacityGap} proved that IA can achieve the
capacity of the real, time-invariant, frequency-flat Gaussian
X-channel to within a constant gap. In fact, for a system designer,
the real concerns are outage probability, ergodic rate and symbol
error rate (SER) at an arbitrary SNR on the time-varying fading
channel. To the best of the authors' knowledge, there is no
literature in solving this problem.

On the other hand, channel state information (CSI) at the
transmitters have a great impact on the performance of IA. Most
previous work analyze the performance assuming that perfect CSI is
available. However, it is difficult to obtain perfect CSI in
practical systems, especially the CSI of the interference channels.
A common way to solve this problem is to exchange CSI between the
transmitters, e.g. CSI is conveyed between the base stations via a
backhaul link in a multi-cell MIMO system. The issues related to CSI
conveyance for IA, including channel estimation, limited CSI
feedback and cooperation, had been discussed in
\cite{LimitedFeedback0}. For interference channels, there are two
CSI exchange modes, namely digital \cite{Digital} and analog
\cite{Analog} transmissions. Digital mode is more popular in IA, due
to its flexibility by adjusting the size of quantization codebook
\cite{LimitedFeedback1}-\cite{LimitedFeedback3}. However,
considering limited feedback bandwidth, the transmitters can only
have partial and imperfect CSI regardless digital or analog feedback
mode. This results in imperfect IA, and hence complicate the
performance analysis.

Existing literature \cite{LimitedFeedback4} showed that even with
limited CSI feedback, the full DOFs of the interference channel can
still be achieved. The average residual interference caused by
imperfect IA in terms of CSI exchange amount was analyzed in
\cite{ResidualInt}, when there is only one data stream for each
transmitter-receiver pair. Then, it was generalized to the case of
multiple data streams \cite{LimitedFeedback2}, where the performance
loss resulting from imperfect CSI was investigated and the
corresponding performance optimization scheme was given. The maximum
DOF and an upper bound on the sum-rate performance loss due to
limited differential feedback rate were given in
\cite{DifferentialFeedback}. A similar case over a two-cell
interfering MIMO-MAC was considered in \cite{Rateloss1}. Moreover,
the impact of feedback rate on the total rate loss of IA for a MIMO
interference channel was analyzed in \cite{Rateloss2}, while a
feedback allocation scheme was proposed to maximize the rate.
Furthermore, the achievable rate of IA in presence of imperfect CSI
was investigated in \cite{MyIA}, and several optimization schemes
were proposed to maximize the rate afterwards. However, a
comprehensive performance analysis of outage probability, ergodic
rate and SER for IA under imperfect CSI, is still unsolved.

\subsection{Main Contributions}
Motivated by the above observations, this paper gives a
comprehensive performance analysis of IA under imperfect CSI over a
general MIMO interference channel. To be precise, each communication
pair undergoes different propagation environment and has distinct
number of data streams. We reveal the accurate relation between the
performances and the amount of available CSI. The major
contributions of this paper can be summarized as follows:
\begin{enumerate}

\item We present a performance analysis framework for IA under
imperfect CSI over a general MIMO interference channel, and derive
the closed-form expressions of outage probability, ergodic rate and
SER in terms of the amount of CSI exchanged and other system
parameters.

\item We analyze the performance loss due to imperfect CSI, and
reveal the impact of CSI accuracy on the system performance.

\item We obtain clear insights on the system performance, which provide
useful guidelines for the system designer.
%\begin{enumerate}
%
%\item Incomplete CSI will result in imperfect IA, and hence there
%exists performance saturation for outage probability, ergodic rate
%and SER. In order to improve the performance continuously as SNR
%increases, the CSI exchange amount should be added logarithmically
%proportional to transmit power, and linearly proportional to the
%number of antennas.
%
%\item CSI is useful for performance improvement at low SNR, so CSI
%exchange is needless under this condition.
%
%\item It is optimal to select single data stream for each link at
%high SNR in the sense of maximizing the ergodic sum rate, while it
%makes sense to use the maximum feasible number of data streams at
%low SNR.
%\end{enumerate}

\end{enumerate}

\subsection{Paper Organization}
The rest of this paper is organized as follows: Section II gives a
brief introduction of the considered MIMO interference network
employing IA under imperfect CSI. Section III focuses on the
performance analysis and derive the closed-form expressions of
outage probability, ergodic rate and SER. Section IV investigates
the performance loss caused by imperfect CSI and get some useful
optimization guidelines. Section V presents several numerical
results to validate the theoretical claims, and finally Section VI
concludes the whole paper.

\emph{Notations}: We use bold upper (lower) letters to denote
matrices (column vectors), $(\cdot)^H$ to denote conjugate
transpose, $E[\cdot]$ to denote expectation, $\|\cdot\|$ to denote
the $L_2$ norm of a vector, $|\cdot|$ to denote the absolute value,
$(a)^{+}$ to denote $\max(a,0)$, $\lceil a\rceil$ to denote the
smallest integer not less than $a$, $\lfloor a\rfloor$ to denote the
largest integer not greater than $a$, and $\stackrel{d}{=}$ to
denote the equality in distribution. The acronym i.i.d. means
``independent and identically distributed", pdf means ``probability
density function" and cdf means ``cumulative distribution function".

\section{System Model}
\begin{figure}[h] \centering
\includegraphics [width=0.45\textwidth] {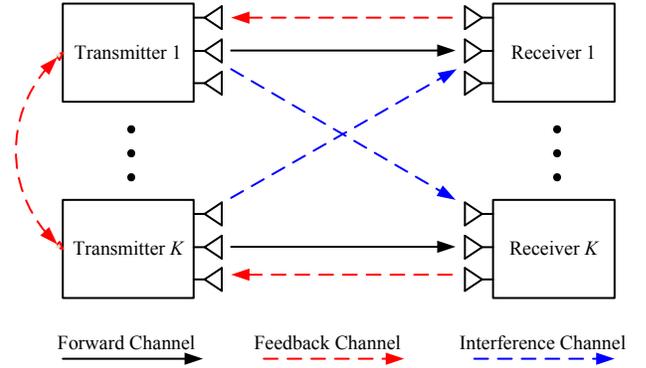}
\caption {A diagram of interference alignment with imperfect CSI
over a MIMO interference channel.} \label{Fig1}
\end{figure}

We consider a MIMO interference network including $K$
transmitter-receiver pairs, as shown in Fig.\ref{Fig1}. For
analytical convenience, we assume an interference network where all
the transmitters and receivers are equipped with $N_t$ and $N_r$
antennas, respectively. Transmitter $k$ sends message(s) to its
intended receiver $k$, while it also creates interference to other
$K-1$ unintended receivers. We use
$\alpha_{k,i}^{1/2}\textbf{H}_{k,i}$ to denote the MIMO channel from
transmitter $i$ to receiver $k$, where $\alpha_{k,i}$ represents the
path loss and $\textbf{H}_{k,i}$ is the $N_r\times N_t$ channel
small scale fading matrix with i.i.d. zero mean and unit variance
complex Gaussian entries. Transmitter $k$ has $d_k$ independent data
streams to be transmitted. It is worth pointing out that due to the
limitation of spatial DOF, the values of $d_k$s must fulfill the
feasibility conditions of IA \cite{Feasibility1}
\cite{Feasibility2}. In what follows, we assume IA is feasible by
choosing $d_k$s carefully. Thus, the received signal at receiver $k$
can be expressed as
\begin{eqnarray}
\textbf{y}_k
&=&\sum\limits_{i=1}^{K}\sqrt{\frac{P\alpha_{k,i}}{d_i}}\textbf{H}_{k,i}\sum\limits_{l=1}^{d_i}\textbf{w}_{i,l}s_{i,l}+\textbf{n}_{k},\label{eqn1}
\end{eqnarray}
where $\textbf{y}_k$ is the $N_r$ dimensional received signal
vector, $\textbf{n}_k$ is the additive Gaussian white noise with
zero mean and variance matrix $\sigma^2\textbf{I}_{N_r}$, $s_{i,l}$
denotes the $l$th normalized data stream from transmitter $i$, and
$\textbf{w}_{i,l}$ is the corresponding $N_t$ dimensional
beamforming vector. $P$ is the total transmit power at each
transmitter, which is equally allocated to the data streams.
Receiver $k$ uses the received vector $\textbf{v}_{k,j}$ of unit
norm to detect its $j$th data stream, which is given by
\begin{eqnarray}
\hat{s}_{k,j}&=&\textbf{v}_{k,j}^{H}\textbf{y}_k\nonumber\\
&=&\sqrt{\frac{P\alpha_{k,k}}{d_k}}\textbf{v}_{k,j}^{H}\textbf{H}_{k,k}\textbf{w}_{k,j}s_{k,j}
+\sqrt{\frac{P\alpha_{k,k}}{d_k}}\sum\limits_{l=1,l\neq
j}^{d_k}\nonumber\\
&&\textbf{v}_{k,j}^{H}\textbf{H}_{k,k}\textbf{w}_{k,l}s_{k,l}
+\sum\limits_{i=1,i\neq
k}^{K}\sqrt{\frac{P\alpha_{k,i}}{d_i}}\sum\limits_{l=1}^{d_i}\nonumber\\
&&\textbf{v}_{k,j}^{H}\textbf{H}_{k,i}\textbf{w}_{i,l}s_{i,l}
+\textbf{v}_{k,j}^{H}\textbf{n}_{k},\label{eqn2}
\end{eqnarray}
where the first term at the right side of (\ref{eqn2}) is the
desired signal, the second one is the inter-stream interference
caused by the same transmitter, and the third one is the inter-link
interference resulting from other transmitters. In order to mitigate
these interferences and improve the performance, IA is performed
accordingly. If perfect CSI is available at all nodes, according to
the principle of IA, we have
\begin{equation}
\textbf{v}_{k,j}^{H}\textbf{H}_{k,k}\textbf{w}_{k,l}=0,\quad l\neq
j, \forall k\in[1,K], \forall l,j\in[1,d_k].\label{eqn3}
\end{equation}
and
\begin{eqnarray}
\textbf{v}_{k,j}^{H}\textbf{H}_{k,i}\textbf{w}_{i,l}=0,\quad i\neq
k, \forall k,i\in[1,K], \nonumber\\ \forall l\in[1,d_i], \forall
j\in[1,d_k].\label{eqn4}
\end{eqnarray}
In brief, inter-stream and inter-link interferences will be canceled
completely if perfect CSI is available. However, in frequency
division duplex (FDD) systems, it is difficult for the transmitters
to obtain perfect CSI, especially the interference CSI. Similar to
previous analogous works, this paper adopts quantization codebook
based CSI conveyance scheme to inform the transmitters the
interference CSI. Specifically, receiver $k$ first quantizes the
channel direction vectors $\tilde{\textbf{h}}_{k,i},\forall
i\in[1,K]$ with different codebooks, where
$\tilde{\textbf{h}}_{k,i}=\frac{{\textbf{h}}_{k,i}}{\|{\textbf{h}}_{k,i}\|}$
and $\textbf{h}_{k,i}=\textmd{vec}(\textbf{H}_{k,i})$ is the
vectorization of $\textbf{H}_{k,i}$. Then, the quantized CSI is
conveyed to the corresponding transmitter. Due to limited feedback,
the transmitters have imperfect CSI $\hat{\textbf{h}}_{k,i}$, and
the relation between the perfect and the imperfect CSI can be
expressed as \cite{RVQ}
\begin{eqnarray}
\tilde{\textbf{h}}_{k,i}=\sqrt{1-\rho_{k,i}}\hat{\textbf{h}}_{k,i}+\sqrt{\rho_{k,i}}\textbf{e}_{k,i},\label{eqn5}
\end{eqnarray}
where $\textbf{e}_{k,i}$ is the quantization error vector with
i.i.d. zero mean and unit variance complex Gaussian entries, and is
independent of $\hat{\textbf{h}}_{k,i}$. $\rho_{k,i}$, scaling from
0 to 1, indicates the CSI accuracy. If $\rho_{k,i}=0$, then
transmitter $i$ has perfect CSI. Intuitively, $\rho_{k,i}$ is
related to the codebook size $2^{B_{k,i}}$ or the CSI exchange
amount $B_{k,i}$, and can be approximated as
$2^{-\frac{B_{k,i}}{N_tN_r-1}}$ \cite{RVQ}. If the CSI exchange
amount tends to infinity, the transmitter can obtain perfect CSI.

\section{Performance Analysis}
In this section, we set off the performance analysis for IA in a
MIMO interference network with imperfect CSI. We put the focus on
three practical performance metrics, namely outage probability,
ergodic rate and SER. Note that we take the $j$th data stream of the
$k$th pair as an example, but the analysis is applicable to all data
streams.

\subsection{Outage Probability}
Although IA is applied, (\ref{eqn3}) and (\ref{eqn4}) do not hold
true any more due to limited CSI exchange. This results in residual
interference, which is also called interference leakage. The signal
to interference and noise ratio (SINR) related to the $j$th data
stream of the $k$th transmitter-receiver pair can be expressed as
\begin{eqnarray}
\gamma_{k,j}&=&\frac{P\alpha_{k,k}}{d_k}\left|\textbf{v}_{k,j}^{H}\textbf{H}_{k,k}\textbf{w}_{k,j}\right|^2\bigg/\bigg(\frac{P\alpha_{k,k}}{d_k}\sum\limits_{l=1,l\neq
j}^{d_k}\nonumber\\
&&\left|\textbf{v}_{k,j}^{H}\textbf{H}_{k,k}\textbf{w}_{k,l}\right|^2
+\sum\limits_{i=1,i\neq
k}^{K}\frac{P\alpha_{k,i}}{d_i}\sum\limits_{l=1}^{d_i}\left|\textbf{v}_{k,j}^{H}\textbf{H}_{k,i}\textbf{w}_{i,l}\right|^2\nonumber\\
&&+\sigma^2\bigg)\nonumber
\end{eqnarray}
\begin{eqnarray}
&=&\frac{\kappa_{k,k}\left|\textbf{h}_{k,k}^H\textbf{T}_{j,j}^{(k,k)}\right|^2}{I_{k,j}+1},\label{eqn6}
\end{eqnarray}
where $\kappa_{k,k}=\frac{P\alpha_{k,k}}{d_k\sigma^2}$,
$\textbf{T}_{j,j}^{(k,k)}=\textbf{w}_{k,i}\otimes\textbf{v}_{k,j}^H$,
$\otimes$ represents the Kronecker product, and $I_{k,j}$ is the
normalized total residual interference to the $j$th data stream of
the $k$th pair. Through IA, the residual interference can be
expressed as
\begin{eqnarray}
I_{k,j}&=&\kappa_{k,k}\sum\limits_{l=1,l\neq
j}^{d_k}\left|\textbf{v}_{k,j}^{H}\textbf{H}_{k,k}\textbf{w}_{k,l}\right|^2\nonumber\\
&&+\sum\limits_{i=1,i\neq
k}^{K}\kappa_{k,i}\sum\limits_{l=1}^{d_i}\left|\textbf{v}_{k,j}^{H}\textbf{H}_{k,i}\textbf{w}_{i,l}\right|^2\nonumber\\
&=&\kappa_{k,k}\sum\limits_{l=1,l\neq
j}^{d_k}\left|\textbf{h}_{k,k}^H\textbf{T}_{j,l}^{(k,k)}\right|^2\nonumber\\&&+\sum\limits_{i=1,i\neq
k}^{K}\kappa_{k,i}\sum\limits_{l=1}^{d_i}\left|\textbf{h}_{k,i}^H\textbf{T}_{j,l}^{(k,i)}\right|^2\nonumber\\
&=&\kappa_{k,k}\rho_{k,k}\|\textbf{h}_{k,k}\|^2\sum\limits_{l=1,l\neq
j}^{d_k}\left|\textbf{e}_{k,k}^H\textbf{T}_{j,l}^{(k,k)}\right|^2\nonumber\\
&&+\sum\limits_{i=1,i\neq
k}^{K}\kappa_{k,i}\rho_{k,i}\|\textbf{h}_{k,i}\|^2\sum\limits_{l=1}^{d_i}\left|\textbf{e}_{k,i}^H\textbf{T}_{j,l}^{(k,i)}\right|^2,\label{eqn7}
\end{eqnarray}
where (\ref{eqn7}) holds true since only partial interference is
canceled based on the IA principle (\ref{eqn3}) and (\ref{eqn4}) due
to imperfect CSI. Then, the outage probability is given by
\begin{eqnarray}
P_{k,j}^{out}&=&P_r(\gamma_{k,j}\leq\gamma_{th})\nonumber\\
&=&\int_0^{\infty}F\left(\gamma_{th}(x+1)\right)g(x)dx,\label{eqn8}
\end{eqnarray}
where $\gamma_{th}$ is the SINR threshold, $F_{S}(x)$ is the cdf of
the desired signal quality
$\kappa_{k,k}\left|\textbf{h}_{k,k}^H\textbf{T}_{j,j}^{(k,k)}\right|^2$,
and $g(x)$ is the pdf of the residual interference $I_{k,j}$. First,
we check the distribution of the desired sinal quality. Since
$\textbf{T}_{j,j}^{(k,k)}$ is designed independently of
$\textbf{h}_{k,k}$ according to the principle of IA,
$\left|\textbf{h}_{k,k}^H\textbf{T}_{j,j}^{(k,k)}\right|^2$ is
$\chi^2(2)$ distributed, so the cdf of
$\kappa_{k,k}\left|\textbf{h}_{k,k}^H\textbf{T}_{j,j}^{(k,k)}\right|^2$
can be expressed as
\begin{eqnarray}
F(x)=1-\exp\left(-\frac{x}{\kappa_{k,k}}\right).\label{eqn9}
\end{eqnarray}
Then, we analyze the distribution of the residual interference.
According to the theory of quantization cell approximation
\cite{RVQ}, $\rho_{k,i}\|\textbf{h}_{k,i}\|^2$ is
$\Gamma(N_tN_r-1,2^{-\frac{B_{k,i}}{N_tN_r-1}})$ distributed.
Moreover, $|\textbf{e}_{k,i}^H\textbf{T}_{j,l}^{(k,i)}|^2$ for
$i=1,\cdots,K$ are i.i.d. $\beta(1,N_tN_r-2)$ distributed, since
$\textbf{T}_{j,l}^{(k,i)}$ of unit norm is independent of
$\textbf{e}_{k,i}$. For the product of a
$\Gamma(N_tN_r-1,2^{-\frac{B_{k,i}}{N_tN_r-1}})$ distributed random
variable and a $\beta(1,N_tN_r-2)$ distributed random variable, it
is equal to $2^{-\frac{B_{k,i}}{N_tN_r-1}}\chi^2(2)$ in distribution
\cite{QCA}. Based on the fact that the sum of $M$ i.i.d. $\chi^2(2)$
distributed random variables is $\chi^2(2M)$ distributed,
$\rho_{k,i}\sum\limits_{l=1}^{d_i}\|\textbf{h}_{k,i}\|^2
\left|\textbf{s}_{k,i}^H\textbf{T}_{j,l}^{(k,i)}\right|^2$ is
$2^{-\frac{B_{k,i}}{N_tN_r-1}}\chi^2(2d_i)$ distributed. Hence,
$I_{k,j}$ is a nested finite weighted sum of $K$ Erlang pdfs, whose
pdf is given by \cite{PDF}
\begin{eqnarray}
g(x)&=&\sum\limits_{i=1}^{K}\sum\limits_{t=1}^{\eta_{k,i}}
\Xi_{K}\Bigg(i,t,\{\eta_{k,q}\}_{q=1}^{K},\left\{\frac{\varrho_{k,q}}{\eta_{k,q}}\right\}_{q=1}^{K},\nonumber\\&&\{l_{k,q}\}_{q=1}^{K-2}\Bigg)
v\left(x,t,\frac{\varrho_{k,i}}{\eta_{k,i}}\right),\label{eqn10}
\end{eqnarray}
where $\eta_{k,i}=d_i$,
$\varrho_{k,i}=\kappa_{k,i}2^{-\frac{B_{k,i}}{N_tN_r-1}}$ for $i\neq
k$, $\eta_{k,k}=d_k-1$,
$\varrho_{k,k}=\kappa_{k,k}2^{-\frac{B_{k,k}}{N_tN_r-1}}$,
$v(x,y,z)=\frac{x^{y-1}\exp(-x/z)}{z^{y}\Gamma(y)},\forall i$, and
$\{x_i\}_{i=1}^{N}$ denotes the set of $x_i$, $i=1,\cdots,N$. The
weights $\Xi_{K}$ are defined as
\begin{eqnarray}
\Xi_{K}\!\!\!\!\!\!\!&&\!\!\!\!\!\!\!\left(i,t,\{\eta_{k,q}\}_{q=1}^{K},\{\varrho_{k,q}\}_{q=1}^{K},\{l_{k,q}\}_{q=1}^{K-2}\right)\nonumber\\
&=&\sum\limits_{l_{k,1}=t}^{\eta_{k,i}}\sum\limits_{l_{k,2}=t}^{l_{k,1}}\cdots\sum\limits_{l_{k,K-2}=t}^{l_{k,K-3}}\bigg[\frac{(-1)^{T_k-\eta_{k,i}}\varrho_{k,i}^{t}}{\prod_{h=1}^{K}\varrho_{k,h}^{\eta_{k,h}}}\nonumber\\
&\times&\frac{\Gamma(\eta_{k,i}+\eta_{k,1+\textmd{U}(1-i)}-l_{k,1})}{\Gamma(\eta_{k,1+\textmd{U}(1-i)})\Gamma(\eta_{k,i}-l_{k,1}+1)}\nonumber\\
&\times&\left(\frac{1}{\varrho_{k,i}}-\frac{1}{\varrho_{k,1+\textmd{U}(1-i)}}\right)^{l_{k,1}-\eta_{k,i}-\eta_{k,1+\textmd{U}(1-i)}}\nonumber\\
&\times&\frac{\Gamma(l_{k,K-2}+\eta_{k,K-1+\textmd{U}(K-1-i)}-t)}{\Gamma(\eta_{k,L-1+\textmd{U}(K-1-i)})\Gamma(l_{k,K-2}-t+1)}\nonumber\\
&\times&\left(\frac{1}{\varrho_{k,i}}-\frac{1}{\varrho_{k,K-1+\textmd{U}(K-1-i)}}\right)^{t-l_{k,K-2}-\eta_{k,K-1+\textmd{U}(K-1-i)}}\nonumber\\
&\times&\prod_{s=1}^{K-3}\frac{\Gamma(l_{k,s}+\eta_{k,s+1+\textmd{U}(s+1-i)}-l_{k,s+1})}{\Gamma(\eta_{k,s+1+\textmd{U}(s+1-i)})\Gamma(l_{k,s}-l_{k,s+1}+1)}\nonumber\\
&\times&\left(\frac{1}{\varrho_{k,i}}-\frac{1}{\varrho_{k,s+1+\textmd{U}(s+1-i)}}\right)^{l_{k,s+1}-l_{k,s}-\eta_{k,s+1+\textmd{U}(s+1-i)}}\bigg],\nonumber\\\label{eqn11}
\end{eqnarray}
where $T_k=\sum_{i=1}^{K}\eta_{k,i}$ and $\textmd{U}(x)$ is the
well-known unit step function defined as $\textmd{U}(x\geq0)=1$ and
zero otherwise. Note that the weights $\Xi_K$ are constant when
given $\eta_{k,i}$ and $\varrho_{k,i}$. Substituting (\ref{eqn9})
and (\ref{eqn11}) into (\ref{eqn8}), we have
\begin{eqnarray}
P_{k,j}^{out}&=&1-\sum\limits_{i=1}^{K}\sum\limits_{t=1}^{\eta_{k,i}}\Xi_{K}\Bigg(i,t,\{\eta_{k,q}\}_{q=1}^{K},\left\{\frac{\varrho_{k,q}}{\eta_{k,q}}\right\}_{q=1}^{K},\nonumber\\&&\{l_{k,q}\}_{q=1}^{K-2}\Bigg)
\int_0^{\infty}\exp\left(-\frac{\gamma_{th}}{\kappa_{k,k}}(x+1)\right)\nonumber\\
&\times&\frac{x^{t-1}\exp(-x/(\varrho_{k,q}/\eta_{k,q}))}{(\varrho_{k,q}/\eta_{k,q})^t\Gamma(t)}dx\nonumber\\
&=&1-\exp\left(-\frac{\gamma_{th}}{\kappa_{k,k}}\right)\sum\limits_{i=1}^{K}\sum\limits_{t=1}^{\eta_{k,i}}\Xi_{K}\Bigg(i,t,\{\eta_{k,q}\}_{q=1}^{K},\nonumber\\
&&\left\{\frac{\varrho_{k,q}}{\eta_{k,q}}\right\}_{q=1}^{K},\{l_{k,q}\}_{q=1}^{K-2}\Bigg)
\left(1+\frac{\varrho_{k,q}\gamma_{th}}{\kappa_{k,k}\eta_{k,q}}\right)^{-t}.\nonumber\\\label{eqn12}
\end{eqnarray}

Thus, we derive the outage probability for the $j$th data stream of
the $k$th pair as a function of the number of data streams $d_k$,
the amount of CSI exchange $B_{k,i}$, and the channel condition
$\kappa_{k,j}$. Note that if each pair has only one data stream, the
result can be further simplified. In this case, there is no
inter-link interference, and $I_{k,j}$ is a weighted sum of $K-1$
$\chi^2(2)$ pdfs, whose pdf is given by \cite{SumofExponential}
\begin{eqnarray}
p_I(x)=\left[\prod\limits_{t=1,t\neq
k}^K\lambda_{k,t}\right]\sum\limits_{i=1,i\neq
k}^K\frac{\exp(-\lambda_{k,i}x)}{\prod\limits_{l=1,l\neq k,l\neq
i}^K(\lambda_{k,l}-\lambda_{k,i})},\label{eqn13}
\end{eqnarray}
where $\lambda_{k,i}=\frac{1}{\varrho_{k,i}}$. Then, submitting
(\ref{eqn9}) and (\ref{eqn11}) into (\ref{eqn8}), the outage
probability in the case of single data stream for each pair is given
by
\begin{eqnarray}
P_{k,j}^{out}&=&1-\left[\prod\limits_{t=1,t\neq
k}^K\lambda_{k,t}\right]\nonumber\\
&&\times\sum\limits_{i=1,i\neq
k}^K\int_0^{\infty}\frac{\exp\left(-\frac{\gamma_{k,j}}{\kappa_{k,k}}(x+1)\right)\exp(-\lambda_{k,i}x)}{\prod\limits_{l=1,l\neq
k,l\neq
i}^K(\lambda_{k,l}-\lambda_{k,i})}dx\nonumber\\
&=&1-\exp\left(-\frac{\gamma_{k,j}}{\kappa_{k,k}}\right)\left[\prod\limits_{t=1,t\neq
k}^K\lambda_{k,t}\right]\nonumber\\
&&\times\sum\limits_{i=1,i\neq
k}^K\frac{\left(\frac{\gamma_{th}}{\kappa_{k,k}}+\lambda_{k,i}\right)^{-1}}{\prod\limits_{l=1,l\neq
k,l\neq i}^K(\lambda_{k,l}-\lambda_{k,i})}\nonumber\\
&=&1-\exp\left(-\frac{\gamma_{k,j}}{\kappa_{k,k}}\right)\left[\prod\limits_{t=1,t\neq
k}^K\lambda_{k,t}\right]\nonumber\\
&&\times\sum\limits_{i=1,i\neq
k}^K\frac{\left(1+\frac{\varrho_{k,i}}{\kappa_{k,k}}\gamma_{th}\right)^{-1}}{\lambda_{k,i}\prod\limits_{l=1,l\neq
k,l\neq i}^K(\lambda_{k,l}-\lambda_{k,i})}.\label{eqn14}
\end{eqnarray}
Note that (\ref{eqn14}) is similar to (\ref{eqn12}) by letting all
$d_i=1$. Therefore, given the channel condition $\kappa_{k,i}$ and
amount of CSI exchange $B_{k,i}$, the outage probability can be
computed according to (\ref{eqn12}) if an arbitrary $d_i$ (namely
$\eta_{k,i}$) $>1$, and (\ref{eqn14}) for all $d_i=1$.

\emph{Remark}: It is found that the outage probability in
(\ref{eqn12}) is independent to the index of data stream $j$, this
is because different data streams of the same pair have the same
signal quality, and undergo the same interference in the statistical
sense.

\subsection{Ergodic Rate}
For the $j$th data stream of the $k$th pair, the ergodic rate based
on IA under imperfect CSI can be expressed as
\begin{eqnarray}
R_{k,j}&=&E[\log_2(1+\gamma_{k,j})]\nonumber\\
&=&\frac{1}{\ln(2)}\bigg(E\left[\ln\left(\kappa_{k,k}\left|\textbf{h}_{k,k}^H\textbf{T}_{j,j}^{(k,k)}\right|^2+I_{k,j}+1\right)\right]\nonumber\\
&&-E\left[\ln\left(I_{k,j}+1\right)\right]\bigg).\label{eqn15}
\end{eqnarray}
For analytical convenience, we use $W_1$ and $W_2$ to denote the
first and the second expectation terms in the bracket at the right
side of (\ref{eqn15}). As discussed above,
$\left|\textbf{h}_{k,k}^H\textbf{T}_{j,j}^{(k,k)}\right|^2$ is
$\chi^2(2)$ distributed, so
$\kappa_{k,k}\left|\textbf{h}_{k,k}^H\textbf{T}_{j,j}^{(k,k)}\right|^2+I_{k,j}$
is a nested finite weighted sum of $K+1$ Erlang pdfs, whose pdf is
given by \cite{PDF}
\begin{eqnarray}
q(x)&=&\sum\limits_{i=1}^{L}\sum\limits_{t=1}^{\omega_{k,i}}\Xi_{L}\Bigg(i,t,\{\omega_{k,q}\}_{q=1}^{L},\left\{\frac{\varrho_{k,q}}{\omega_{k,q}}\right\}_{q=1}^{L},\nonumber\\&&\{l_{k,q}\}_{q=1}^{L-2}\Bigg)
v\left(x,t,\frac{\varrho_{k,i}}{\omega_{k,i}}\right),\label{eqn16}
\end{eqnarray}
where $L=K+1$, $\omega_{k,i}=d_i,\forall i\neq k$,
$\omega_{k,k}=d_k-1$, $\omega_{k,L}=1$ and
$\varrho_{k,L}=\kappa_{k,k}$. Hence, $W_1$ can be computed as
\begin{eqnarray}
W_1&=&\sum\limits_{i=1}^{L}\sum\limits_{t=1}^{\omega_{k,i}}\Xi_{L}\left(i,t,\{\omega_{k,q}\}_{q=1}^{L},\left\{\frac{\varrho_{k,q}}{\omega_{k,q}}\right\}_{q=1}^{L},\{l_{k,q}\}_{q=1}^{L-2}\right)\nonumber\\
&&\times\int_0^{\infty}\ln(x+1)\frac{x^{t-1}\exp\left(-\frac{x}{(\varrho_{k,q}/\omega_{k,q})}\right)}{(\varrho_{k,q}/\omega_{k,q})^t\Gamma(t)}dx\nonumber\\
&=&\sum\limits_{i=1}^{L}\sum\limits_{t=1}^{\omega_{k,i}}\Xi_{L}\left(i,t,\{\omega_{k,q}\}_{q=1}^{L},\left\{\frac{\varrho_{k,q}}{\omega_{k,q}}\right\}_{q=1}^{L},\{l_{k,q}\}_{q=1}^{L-2}\right)\nonumber\\
&&\times
Z\left(t,\frac{\varrho_{k,q}}{\omega_{k,q}}\right),\label{eqn17}
\end{eqnarray}
where
\begin{eqnarray}
Z\left(t,\frac{\varrho_{k,q}}{\omega_{k,q}}\right)
&=&\sum\limits_{\vartheta=0}^{t-1}\frac{1}{\Gamma(t-\vartheta)}\bigg((-1)^{t-\vartheta-2}\left(\frac{\omega_{k,q}}{\varrho_{k,q}}\right)^{t-\vartheta-1}\nonumber\\
&&\times\exp\left(\frac{\omega_{k,q}}{\varrho_{k,q}}\right)
\textmd{E}_{\textmd{i}}\left(-\frac{\omega_{k,q}}{\varrho_{k,q}}\right)\nonumber\\
&&+\sum\limits_{\nu=1}^{t-\vartheta-1}\Gamma(\nu)\left(-\frac{\omega_{k,q}}{\varrho_{k,q}}\right)^{t-\vartheta-\nu-1}\bigg),\label{eqn53}
\end{eqnarray}
and
$\textmd{E}_{\textmd{i}}(x)=\int_{-\infty}^{x}\frac{\exp(t)}{t}dt$
is the exponential integral function. (\ref{eqn17}) is obtained
based on [36, Eq.4.337.5]. Similarly, we can get $W_2$ as follows:
\begin{eqnarray}
W_2&=&\sum\limits_{i=1}^{K}\sum\limits_{t=1}^{\eta_{k,i}}\Xi_{K}\Bigg(i,t,\{\eta_{k,q}\}_{q=1}^{K},\left\{\frac{\varrho_{k,q}}{\eta_{k,q}}\right\}_{q=1}^{K},\{l_{k,q}\}_{q=1}^{K-2}\Bigg)\nonumber\\
&&\times\int_0^{\infty}\ln(x+1)\frac{x^{t-1}\exp\left(-\frac{x}{(\varrho_{k,q}/\eta_{k,q})}\right)}{(\varrho_{k,q}/\eta_{k,q})^t\Gamma(t)}dx\nonumber\\
&=&\sum\limits_{i=1}^{K}\sum\limits_{t=1}^{\eta_{k,i}}\Xi_{K}\left(i,t,\{\eta_{k,q}\}_{q=1}^{K},\left\{\frac{\varrho_{k,q}}{\eta_{k,q}}\right\}_{q=1}^{K},\{l_{k,q}\}_{q=1}^{K-2}\right)\nonumber\\
&&\times
Z\left(t,\frac{\varrho_{k,q}}{\eta_{k,q}}\right).\label{eqn18}
\end{eqnarray}
Based on (\ref{eqn17}) and (\ref{eqn18}), the ergodic rate is given
by
\begin{eqnarray}
R_{k,j}&=&\frac{1}{\ln(2)}\bigg(\sum\limits_{i=1}^{L}\sum\limits_{t=1}^{\omega_{k,i}}\Xi_{L}\Bigg(i,t,\{\omega_{k,q}\}_{q=1}^{L},\left\{\frac{\varrho_{k,q}}{\omega_{k,q}}\right\}_{q=1}^{L},\nonumber\\&&\{l_{k,q}\}_{q=1}^{L-2}\Bigg)Z\left(t,\frac{\varrho_{k,q}}{\omega_{k,q}}\right)\nonumber\\
&&-\sum\limits_{i=1}^{K}\sum\limits_{t=1}^{\eta_{k,i}}\Xi_{K}\Bigg(i,t,\{\eta_{k,q}\}_{q=1}^{K},\left\{\frac{\varrho_{k,q}}{\eta_{k,q}}\right\}_{q=1}^{K},\nonumber\\&&\{l_{k,q}\}_{q=1}^{K-2}\Bigg)Z\left(t,\frac{\varrho_{k,q}}{\eta_{k,q}}\right)\bigg).\label{eqn19}
\end{eqnarray}

Additionally, in the case of single data stream for each pair, the
ergodic rate can be computed as
\begin{eqnarray}
R_{k,j}&=&E\left[\log_2\left(1+\frac{\kappa_{k,k}\left|\textbf{h}_{k,k}^H\textbf{T}_{j,j}^{(k,k)}\right|^2}{I_{k,j}+1}\right)\right]\nonumber\\
&=&\frac{1}{\ln(2)}\bigg(E\left[\ln\left(\kappa_{k,k}\left|\textbf{h}_{k,k}^H\textbf{T}_{j,j}^{(k,k)}\right|^2+I_{k,j}+1\right)\right]\nonumber\\
&&-E[\ln(I_{k,j}+1)]\bigg)\nonumber\\
&=&\frac{1}{\ln(2)}\Bigg(\left[\prod\limits_{t=1}^K\lambda_{k,t}\right]\sum\limits_{i=1}^K\frac{\int_0^{\infty}\ln(x+1)\exp(-\lambda_{k,i}x)dx}{\prod\limits_{l=1,l\neq
i}^K(\lambda_{k,l}-\lambda_{k,i})}\nonumber\\
&&-\left[\prod\limits_{t=1,t\neq
k}^K\lambda_{k,t}\right]\sum\limits_{i=1,i\neq
k}^K\frac{\int_0^{\infty}\ln(x+1)\exp(-\lambda_{k,i}x)dx}{\prod\limits_{l=1,l\neq
k,l\neq i}^K(\lambda_{k,l}-\lambda_{k,i})}\Bigg)\nonumber\\
&=&\frac{1}{\ln(2)}\Bigg(\left[\prod\limits_{t=1}^K\lambda_{k,t}\right]\sum\limits_{i=1}^K\frac{\exp(\lambda_{k,i})\textmd{E}_{\textmd{i}}(\lambda_{k,i})}{\lambda_{k,i}\prod\limits_{l=1,l\neq
i}^K(\lambda_{k,l}-\lambda_{k,i})}\nonumber\\
&&-\left[\prod\limits_{t=1,t\neq
k}^K\lambda_{k,t}\right]\sum\limits_{i=1,i\neq
k}^K\frac{\exp(\lambda_{k,i})\textmd{E}_{\textmd{i}}(\lambda_{k,i})}{\lambda_{k,i}\prod\limits_{l=1,l\neq
k,l\neq
i}^K(\lambda_{k,l}-\lambda_{k,i})}\Bigg),\nonumber\\\label{eqn21}
\end{eqnarray}
where (\ref{eqn21}) is obtained based on [36, Eq.4.337.2]. Note that
(\ref{eqn21}) is equivalent to (\ref{eqn19}) because of
$\lambda_{k,i}=\frac{1}{\varrho_{k,i}}$ when $t=1$.

\subsection{SER}
In addition to outage probability and ergodic rate, SER is also an
important performance metric, which indicates the reliability of a
wireless communication system. For a given SINR $\gamma$, the SER is
given by \cite{SER}
\begin{eqnarray}
\textrm{SER}=\left\{\begin{array}{ll}\frac{1}{\pi}\int_0^{\frac{(M-1)\pi}{M}}\exp\left(-\frac{g_{\textrm{PSK}}\gamma}{\sin^2(x)}\right)dx,&\textrm{for
$M$-PSK}\\
\frac{2}{\pi}\frac{M-1}{M}\int_0^{\frac{\pi}{2}}\exp\left(-\frac{g_{\textrm{PAM}}\gamma}{\sin^2(x)}\right)dx,&\textrm{for $M$-PAM}\\
\frac{4}{\pi}\left(1-\frac{1}{\sqrt{M}}\right)\\\bigg[\frac{1}{\sqrt{M}}\int_0^{\frac{\pi}{4}}\exp\left(-\frac{g_{\textrm{QAM}}\gamma}{\sin^2(x)}\right)dx\\+\int_{\frac{\pi}{4}}^{\frac{\pi}{2}}\exp\left(-\frac{g_{\textrm{QAM}}\gamma}{\sin^2(x)}\right)dx\bigg],&\textrm{for
$M$-QAM}
\end{array}\right.\label{eqn22}
\end{eqnarray}
where $M$ is the modulation order,
$g_{\textrm{PSK}}=\sin^2\frac{\pi}{M}$,
$g_{\textrm{PAM}}=\frac{3}{M^2-1}$ and
$g_{\textrm{AQM}}=\frac{3}{2(M-1)}$. For analytical convenience, we
use $G\left(M,a\exp(-b\gamma)\right)$ to denote a general SER
function, where $a$ and $b$ depend on the modulation format. Then,
the average SER of the $j$th data stream of the $k$th pair can be
computed as
\begin{eqnarray}
P_{k,j}^{ser}&=&\int_0^{\infty}\int_0^{\infty}G\left(M,a\exp\left(-b\frac{x}{y+1}\right)\right)\nonumber\\
&&\times f(x)g(y)dxdy\nonumber\\
&=&\int_0^{\infty}G\left(M,\frac{ay+a}{y+b\kappa_{k,k}+1}\right)g(y)dy\label{eqn23}
\end{eqnarray}
\begin{eqnarray}
&=&G\bigg(M,\sum\limits_{i=1}^{K}\sum\limits_{t=1}^{\eta_{k,i}}
\Xi_{K}\Bigg(i,t,\{\eta_{k,q}\}_{q=1}^{K},\left\{\frac{\varrho_{k,q}}{\eta_{k,q}}\right\}_{q=1}^{K},\nonumber\\&&\{l_{k,q}\}_{q=1}^{K-2}\Bigg)\nonumber\\
&&\times a\bigg(1-b\kappa_{k,k}(b\kappa_{k,k}+1)^{t-1}
\exp\left(\frac{(b\kappa_{k,k}+1)\eta_{k,q}}{\varrho_{k,q}}\right)\nonumber\\
&&\times\Gamma\left(1-t,\frac{(b\kappa_{k,k}+1)\eta_{k,q}}{\varrho_{k,q}}\right)\bigg)\bigg),\label{eqn24}
\end{eqnarray}
where
$f(x)=\frac{1}{\kappa_{k,k}}\exp\left(-\frac{x}{\kappa_{k,k}}\right)$
is the pdf of
$\kappa_{k,k}\left|\textbf{h}_{k,k}^H\textbf{T}_{j,j}^{(k,k)}\right|^2$.
(\ref{eqn23}) follows the fact that $G(M,\\a\exp(-b\gamma))$ is a
linear function of $a\exp\left(-b\frac{x}{y+1}\right)$, and
(\ref{eqn24}) is obtained based on [36, Eq.3.383.10].

Similarly, for the case of single data stream, the average SER can
be reduced as
\begin{eqnarray}
P_{k,j}^{ser}&=&\int_0^{\infty}\int_0^{\infty}G\left(M,a\exp\left(-b\frac{x}{y+1}\right)\right)\nonumber\\
&&\times f_S(x)p_I(y)dxdy\nonumber\\
&=&\int_0^{\infty}G\left(M,\frac{ay+a}{y+1+b\kappa_{k,k}}\right)p_I(y)dy\nonumber\\
&=&G\Bigg(M,\left[\prod\limits_{t=1,t\neq
k}^K\lambda_{k,t}\right]\sum\limits_{i=1,i\neq
k}^Ka(1-b\kappa_{k,k}\nonumber\\&&\times\exp\left((b\kappa_{k,k}+1)\lambda_{k,i}\right)
\Gamma(0,(b\kappa_{k,k}+1)\lambda_{k,i}))\nonumber\\
&&/(\lambda_{k,i}\prod\limits_{l=1,l\neq k,l\neq
i}^K(\lambda_{k,l}-\lambda_{k,i}))\Bigg).\label{eqn25}
\end{eqnarray}
Note that (\ref{eqn25}) is also equivalent to (\ref{eqn24}) by
letting $t=\eta_{k,i}=1$ and $\lambda_{k,i}=1/\varrho_{k,i}$.
Submitting (\ref{eqn24}) and (\ref{eqn25}) into (\ref{eqn22}), we
could obtain the expression of SER. As a simple example, for $M$-PSK
modulation, $a=\frac{1}{\pi}$ and
$b=\frac{g_{\textrm{PSK}}}{\sin^2(x)}$, so the average SER is given
by
\begin{eqnarray}
P_{k,j}^{ser}&=&\frac{1}{\pi}\int_0^{\frac{(M-1)\pi}{M}}\sum\limits_{i=1}^{K}\sum\limits_{t=1}^{\eta_{k,i}}\Xi_{K}\Bigg(i,t,\{\eta_{k,q}\}_{q=1}^{K},\nonumber\\&&\left\{\frac{\varrho_{k,q}}{\eta_{k,q}}\right\}_{q=1}^{K},\{l_{k,q}\}_{q=1}^{K-2}\Bigg)\nonumber\\
&&\times\bigg(1-\frac{g_{\textrm{PSK}}\kappa_{k,k}}{\sin^2(x)}\left(\frac{g_{\textrm{PSK}}\kappa_{k,k}+\sin^2(x)}{\sin^2(x)}\right)^{t-1}\nonumber\\
&&\times\exp\left(\frac{(g_{\textrm{PSK}}\kappa_{k,k}+\sin^2(x))\eta_{k,q}}{\sin^2(x)\varrho_{k,q}}\right)\nonumber\\
&&\times\Gamma\left(1-t,\frac{(g_{\textrm{PSK}}+\sin^2(x))\kappa_{k,k}\eta_{k,q}}{\sin^2(x)\varrho_{k,q}}\right)\bigg)dx.\nonumber\\\label{eqn26}
\end{eqnarray}
Given channel condition and amount of CSI exchange, $P_{k,j}^{ser}$
can be computed by numerical integration.

\section{Performance Loss from Imperfect CSI}
Due to limited CSI feedback, there exists a certain performance loss
with respect to the case of perfect CSI due to imperfect IA. In
order to reveal the impact of CSI and obtain insightful guidelines
for system design and performance optimization, we investigate the
performance loss resulting from imperfect CSI from the standpoints
of outage probability, ergodic rate and SER, respectively. For the
convenience of analysis, we use $\Upsilon=\frac{P}{\sigma^2}$ to
denote the transmit SNR.

\subsection{Outage Probability}
If the transmitters have perfect CSI, the interference can be
canceled completely, and then the SINR of the $j$th data stream of
the $k$th pair is transformed as
\begin{eqnarray}
\gamma_{k,j}^{perfect}=\kappa_{k,k}\left|\textbf{h}_{k,k}^H\textbf{T}_{j,j}^{(k,k)}\right|^2.\label{eqn27}
\end{eqnarray}
In this context, the outage probability of the $j$th data stream of
the $k$th pair related to a given SNR threshold $\gamma_{th}$ can be
computed as
\begin{eqnarray}
P_{k,j}^{out,perfect}&=&P_r(\gamma_{k,j}^{perfect}\leq\gamma_{th})\nonumber\\
&=&F(\gamma_{th})\nonumber\\
&=&1-\exp\left(-\frac{\gamma_{th}}{\kappa_{k,k}}\right).\label{eqn28}
\end{eqnarray}
Based on (\ref{eqn12}) and (\ref{eqn28}), the performance loss due
to imperfect CSI is given by:
\begin{eqnarray}
\Delta P_{k,j}^{out}&=&P_{k,j}^{out}-P_{k,j}^{out,full}\nonumber\\
&=&\exp\left(-\frac{\gamma_{th}}{\kappa_{k,k}}\right)\Bigg(1-\sum\limits_{i=1}^{K}\sum\limits_{t=1}^{\eta_{k,i}}\Xi_{K}\Bigg(i,t,\{\eta_{k,q}\}_{q=1}^{K},\nonumber\\&&\left\{\frac{\varrho_{k,q}}{\eta_{k,q}}\right\}_{q=1}^{K},\{l_{k,q}\}_{q=1}^{K-2}\Bigg)
\left(1+\frac{\varrho_{k,q}\gamma_{th}}{\kappa_{k,k}\eta_{k,q}}\right)^{-t}\Bigg)\label{eqn35}\\
&=&\exp\left(-\frac{\gamma_{th}}{\Upsilon\alpha_{k,k}}\right)\Bigg(1-\sum\limits_{i=1}^{K}\sum\limits_{t=1}^{\eta_{k,i}}\Xi_{K}\Bigg(i,t,\{\eta_{k,q}\}_{q=1}^{K},\nonumber\\&&\left\{\frac{\varrho_{k,q}}{\eta_{k,q}}\right\}_{q=1}^{K},\{l_{k,q}\}_{q=1}^{K-2}\Bigg)
\left(1+\frac{\alpha_{k,q}\rho_{k,q}\gamma_{th}}{\alpha_{k,k}\eta_{k,q}}\right)^{-t}\Bigg).\nonumber\\\label{eqn29}
\end{eqnarray}
(\ref{eqn29}) provides a general outage probability loss due to
imperfect CSI in terms of CSI accuracy, transmit SNR, and channel
condition. However, it is difficult to get some clear insights from
(\ref{eqn29}). To address this problem, we perform asymptotic
analysis under some extreme scenario. First, if SNR $\Upsilon$ is
low enough, we have the following theorem:

\emph{Theorem 1}: At low SNR, the CSI exchange is useless, and the
performance loss due to imperfect CSI is negligible.

\begin{proof}
The proof is straightforward. When SNR $\Upsilon\rightarrow0$, the
term $\exp\left(-\frac{\gamma_{th}}{\Upsilon\alpha_{k,k}}\right)$ in
(\ref{eqn29}) asymptotically approaches zero, so there is no
performance gap between the two cases, and hence the CSI is useless.
\end{proof}

\emph{Remark}: If SNR is quite low, the interference term becomes
negligible with respect to the noise. So the CSI exchange is not
needed, as well as the IA. In this context, maximum ratio
transmission (MRT) can obtain the maximum DOFs, and hence achieves
the optimal performance.

On the other hand, if SNR $\Upsilon$ is sufficiently high, the
performance loss in terms of outage probability has the following
property:

\emph{Theorem 2}: At high SNR, there is always a performance floor
in terms of outage probability for IA with imperfect CSI. The
performance loss becomes larger as SNR increases, and the maximum
performance loss is given by $\Delta
P_{k,j}^{out,\max}=1-\sum\limits_{i=1}^{K}\sum\limits_{t=1}^{\eta_{k,i}}
\Xi_{K}\bigg(i,t,\{\eta_{k,q}\}_{q=1}^{K},\bigg\{\frac{\varrho_{k,q}}{\eta_{k,q}}\bigg\}_{q=1}^{K},\{l_{k,q}\}_{q=1}^{K-2}\bigg)\\
\left(1+\frac{\alpha_{k,q}\rho_{k,q}\gamma_{th}}{\alpha_{k,k}\eta_{k,q}}\right)^{-t}$.

\begin{proof}
Please refer to Appendix I.
\end{proof}

From Theorem 2, it is known that the performance gap of outage
probability caused by imperfect CSI is an increasing function of
$\Upsilon$ and $\rho_{k,i}$, so the performance gap can be reduced
by adding the amount of CSI exchange. Furthermore, for the sake of
keeping a constant gap as SNR increases, the amount of CSI exchange
should satisfy the following proposition:

\emph{Proposition 1}: In order to avoid the performance floor and
keep a constant gap between imperfect and perfect CSI, the total
amount of CSI exchange from the $k$th receiver should fulfill the
condition of $B_k\sim K(N_tN_r-1)\log_2(\Upsilon)$, where $\sim$
denotes proportional to.

\begin{proof}
Please refer to Appendix II.
\end{proof}

Based on Proposition 1, we further obtain a useful guideline for
performance optimization of IA under imperfect CSI as follows

\emph{Proposition 2}: In order to keep a constant gap of outage
probability between imperfect and perfect CSI, the total amount of
CSI exchange $B$ should be added as the number of antennas
increases.

\begin{proof}
The proof can be dervied from Proposition 1 directly. Because of
$B_k\sim K(N_tN_r-1)\log_2(\Upsilon)$, $B_k$ should be added as the
number of antennas increases, which is also consistent with the
intuition that for a given $B_k$, increasing the number of antennas
results in the reduction of CSI quantization accuracy, and hence
leads to performance degradation.
\end{proof}

\subsection{Ergodic Rate}
In this subsection, we turn the attention to the rate loss caused by
imperfect CSI, which is given by
\begin{eqnarray}
\Delta R_{k,j}&=&E\left[\log_2\left(1+\gamma_{k,j}^{perfect}\right)\right]-E[\log_2\left(1+\gamma_{k,j}\right)]\nonumber\\
&=&\frac{1}{\ln(2)}\left(\exp\left(\frac{1}{\kappa_{k,k}}\right)\textmd{E}_{\textmd{i}}\left(\frac{1}{\kappa_{k,k}}\right)\right)-R_{k,j}.\label{eqn30}
\end{eqnarray}
Similarly, we perform asymptotic analysis on $\Delta R_{k,j}$ to get
some clear insights. First, if SNR $\Upsilon$ is low enough, we have
the following theorem:

\emph{Theorem 3}: At low SNR, the CSI exchange is useless. The
ergodic rates in the cases of both perfect and imperfect CSI
approach 0, so the rate loss is negligible.

\begin{proof}
The proof is intuitive, as SNR tends to 0, the interference term is
negligible with respect to the noise term, then the ergodic rate
with imperfect CSI is equivalent to that with perfect CSI, and hence
the performance gap becomes zero.
\end{proof}

On the other hand, for the high SNR case, the rate loss due to
imperfect CSI has the following theorem:

\emph{Theorem 4}: At high SNR, there is always a performance ceiling
in terms of ergodic rate for IA with imperfect CSI, and the
performance loss with respect to the ideal case of perfect CSI
enlarges logarithmically with the SNR.

\begin{proof}
Please refer to Appendix III.
\end{proof}

As analyzed above, the performance ceiling is a decreasing function
of $\rho_{k,i}$. To keep a constant gap, the amount of CSI exchange
should be added as SNR increases. It is assumed that the CSI
exchange amount is the same and quite large in the high SNR region,
then we have the following proposition:

\emph{Proposition 3}: At high SNR with large amount of CSI exchange,
the performance ceiling is a linear function of the amount of CSI
exchange, and the performance gain by adding additional amount of
CSI exchange is equal to $\frac{1}{N_tN_r-1}$ times the incremental
amount of CSI.

\begin{proof}
Please refer to Appendix IV.
\end{proof}

From Theorem 4, it is known that the ergodic rate with perfect CSI
at high SNR is
$R_{k,j}^{perfect,high}=\frac{\ln(\Upsilon\alpha_{k,k})-C}{\ln(2)}$,
so the performance loss at high SNR when $B$ is sufficiently large
can be approximated as
\begin{eqnarray}
\Delta R_{k,j}^{high,large}&=&\log_2(\Upsilon)-\frac{B}{N_tN_r-1}+\frac{1}{\ln(2)}\sum\limits_{i=1}^{K}\sum\limits_{t=1}^{\eta_{k,i}}\nonumber\\&&\Xi_{K}\left(i,t,\{\eta_{k,q}\}_{q=1}^{K},\left\{\frac{\xi_{k,q}}{\eta_{k,q}}\right\}_{q=1}^{K},\{l_{k,q}\}_{q=1}^{K-2}\right)\nonumber\\
&&\times\left(\psi(t)+\ln(\alpha_{k,q})-\ln(\eta_{k,q})\right).\label{eqn31}
\end{eqnarray}
As mentioned in Theorem 4, the performance loss with respect to the
ideal case of perfect CSI enlarges logarithmically with the SNR.
Based on (\ref{eqn31}), we have the following proposition:

\emph{Proposition 4}: At high SNR, in order to keep a constant
performance gap, $B$ should be added proportionally to
$(N_tN_r-1)\log_2(\Upsilon)$.

%Note that for ergodic rate, it is better to maximize the sum rate
%than to optimize a certain data stream. Then, the number of data
%streams for a link has a great impact on the overall performance.
%The large number of data streams can exploit more spatial
%multiplexing gains, but also leads to higher residual interference,
%so it is necessary to select an optimal number of data streams.
%Following the Theorem 3, we can obtain a proposition as below:
%
%\emph{Proposition 3}: At low SNR, it is optimal to select the
%maximum feasible number of data streams.
%
%\begin{proof}
%As Theorem 3 claims, the CSI amount seldom affects the ergodic rate
%of each data stream, since the interference is negligible at very
%low SNR. Then, using more data stream can add the multiplexing
%gains, but hardly degrades the performance of each stream, so it is
%optimal to adopt the maximum feasible number of data streams, which
%must fulfill the feasibility condition of IA \cite{Feasibility1}
%\cite{Feasibility2}.
%\end{proof}
%
%On the contrary, if SNR is sufficiently high, the number of data
%stream should be chosen according to the following proposition:
%
%\emph{Proposition 4}: At high SNR, it is optimal to select single
%data stream for each link in the sense of maximizing the ergodic sum
%rate.
%
%\begin{proof}
%Please refer to Appendix V.
%\end{proof}

\subsection{SER}
Intuitively, imperfect CSI would lead to SER performance degradation
inevitably. In this subsection, we give a quantitative analysis of
the performance loss. Substituting (\ref{eqn27}) into (\ref{eqn22}),
the average SER with perfect CSI can be computed as
\begin{eqnarray}
P_{k,j}^{ser,perfect}&=&\int_{0}^{\infty}G\left(M,a\exp(-bx)\right)\frac{1}{\kappa_{k,k}}\exp\left(-\frac{x}{\kappa_{k,k}}\right)dx\nonumber\\
&=&G\left(M,\frac{a}{1+b\kappa_{k,k}}\right).\label{eqn32}
\end{eqnarray}
Thereby, the performance gap is given by
\begin{eqnarray}
\Delta
P_{k,j}^{ser}&=&G\bigg(M,\sum\limits_{i=1}^{K}\sum\limits_{t=1}^{\eta_{k,i}}
\Xi_{K}\Bigg(i,t,\{\eta_{k,q}\}_{q=1}^{K},\left\{\frac{\varrho_{k,q}}{\eta_{k,q}}\right\}_{q=1}^{K},\nonumber\\&&\{l_{k,q}\}_{q=1}^{K-2}\Bigg)
a\bigg(1-b\kappa_{k,k}(b\kappa_{k,k}+1)^{t-1}\nonumber\\
&&\times\exp\left(\frac{(b\kappa_{k,k}+1)\eta_{k,q}}{\varrho_{k,q}}\right)\nonumber\\
&&\times\Gamma\left(1-t,\frac{(b\kappa_{k,k}+1)\eta_{k,q}}{\varrho_{k,q}}\right)\bigg)-\frac{a}{1+b\kappa_{k,k}}\bigg),\nonumber\\\label{eqn33}
\end{eqnarray}
where (\ref{eqn33}) holds true since $G(x,y)$ is a linear function
of $y$. Similarly, we carry out asymptotic analysis on the
performance loss to get some clear insights. First, if SNR is low
enough, we have the following theorem:

\emph{Theorem 5}: At low SNR, the CSI exchange is useless, the
average SER in the cases of both perfect and imperfect CSI
approaches 0.5, so the performance gap is negligible.

\begin{proof}
The proof is similar to that of Theorem 1.
\end{proof}

Furthermore, when SNR asymptotically approaches infinity, the
performance loss has the following theorem:

\emph{Theorem 6}: At high SNR, there is a SER performance floor for
IA with imperfect CSI. The performance gap becomes larger as SNR
increases.

The floor appears since the SINR is saturated once SNR exceeds a
specific value, so it is necessary to add the amount of CSI exchange
logarithmically proportional to SNR in order to avoid the floor.

\section{Numerical Results}
To verify the performance analysis results for IA with imperfect CSI
in a MIMO interference network, we present several numerical results
under different scenarios. For convenience, we set $N_t=4$, $N_r=2$,
$K=3$, $\sigma^2=1$, $\gamma_{th}=0$ dB, $d_1=d_2=d_3=d=1$ and
$\alpha_{i,j}$ given in Tab.\ref{Tab1} for all simulation scenarios
without explicit explanation. Note that we consider an interference
channel model as shown in Fig.1, the propagation distances for the
desired links are the same, and the distance between the
interference transmitter to the receiver is larger than that between
the desired transmitter to the receiver. Mathematically, we have
$\alpha_{i,i}>\alpha_{i,j}$, if $i\neq j$. In convenience, we
normalize $\alpha_{i,i}$ as 1, so we have $\alpha_{i,j}<1$ for
$i\neq j$. In addition, we use SNR (in dB) to represent
$10\log_{10}\frac{P}{\sigma^2}$, and $B$ (in bit) to denote the same
amount CSI exchange for each pair. Without loss of generality, we
take the performance of the $1$st data stream of $1$st pair as an
example.

\newcommand{\tabincell}[2]{\begin{tabular}{@{}#1@{}}#2\end{tabular}}
\begin{table}\centering
\caption{Parameter Table for $\alpha_{i,j}, \forall i,j\in[1,3]$.}
\label{Tab1}
\begin{tabular*}{4.42cm}{|c|c|c|c|}\hline
\backslashbox{i}{j}& 1 & 2 & 3\\
\hline 1 & 1.000 & 0.050 & 0.005 \\
\hline 2 & 0.055 & 1.000 & 0.045 \\
\hline 3 & 0.004 & 0.060 & 1.000 \\
\hline
\end{tabular*}
\end{table}

\begin{figure}[h] \centering
\includegraphics [width=0.5\textwidth] {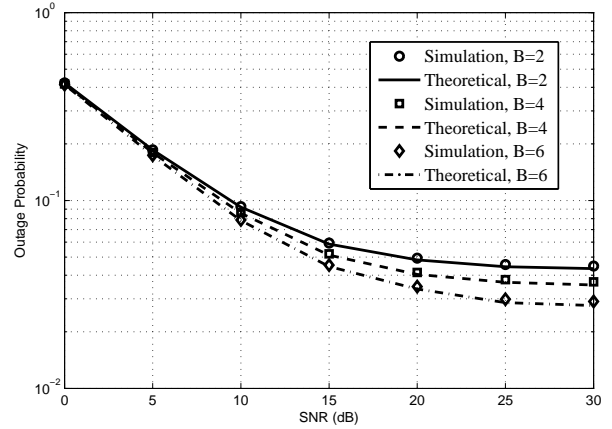}
\caption {Theoretical and simulation outage probability with
different CSI exchange amount.} \label{Fig2}
\end{figure}

\begin{figure}[h] \centering
\includegraphics [width=0.5\textwidth] {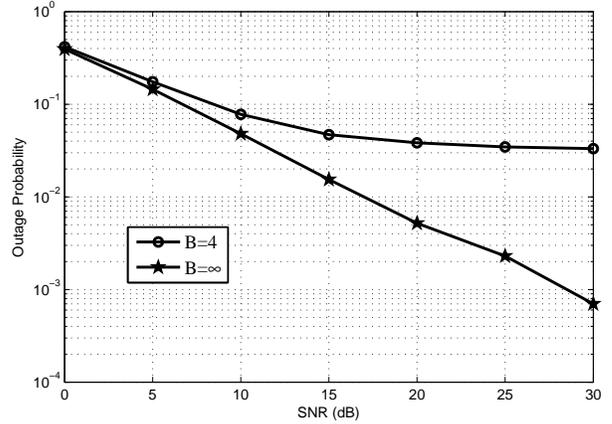}
\caption {Outage probability loss due to imperfect CSI.}
\label{Fig3}
\end{figure}

Firstly, we investigate the outage probability of IA with imperfect
CSI. As seen in Fig.\ref{Fig2}, the theoretical results are quite
consistent with the simulation results throughout the whole SNR
region under different amount of CSI exchange $B$, which validates
the high accuracy of our analysis. It is found that at low SNR, the
outage probabilities with different $B$ are nearly the same, which
reconfirms the claim of Theorem 1 that the CSI exchange in the low
SNR region is useless. With the increase of the SNR, as Theorem 2
claims, there always exist a performance floor, which decreases as
$B$ adds. In addition, as shown in Fig.\ref{Fig3}, the performance
loss due to imperfect CSI becomes larger as SNR increases. In order
to keep a constant gap, $B$ should be added logarithmically and
linearly proportional to the SNR and the number of antennas
respectively, as proved in Proposition 1 and 2.

\begin{figure}[h] \centering
\includegraphics [width=0.5\textwidth] {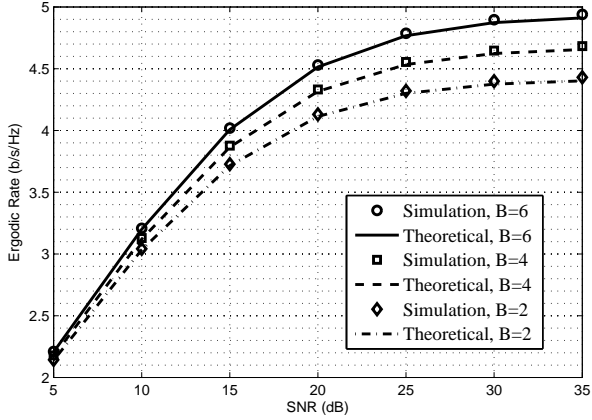}
\caption {Theoretical and simulation ergodic rate performance with
different CSI exchange amount.} \label{Fig4}
\end{figure}

\begin{figure}[h] \centering
\includegraphics [width=0.5\textwidth] {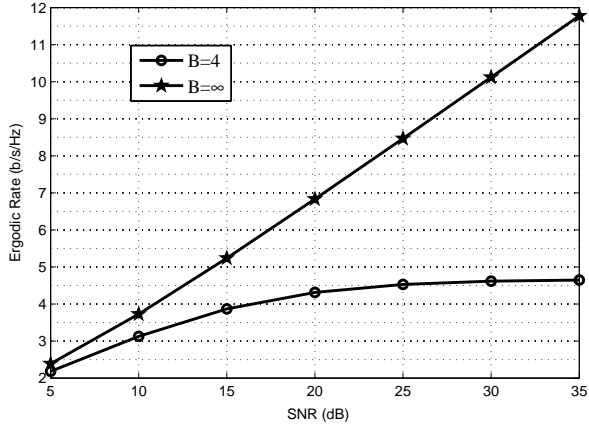}
\caption {Ergodic rate performance loss due to imperfect CSI.}
\label{Fig5}
\end{figure}

\begin{figure}[h] \centering
\includegraphics [width=0.5\textwidth] {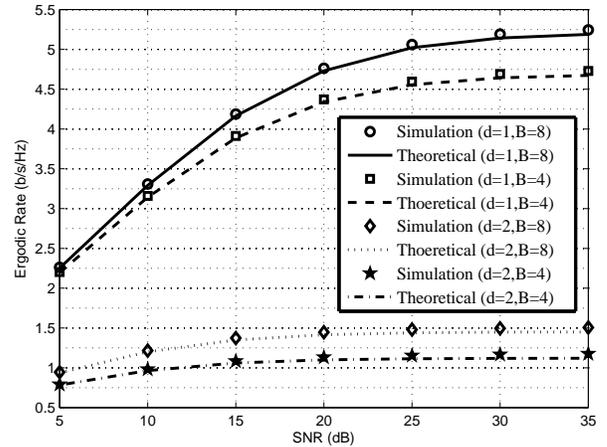}
\caption {Ergodic rate performance with different numbers of data
streams.} \label{Fig6}
\end{figure}

\begin{figure}[h] \centering
\includegraphics [width=0.5\textwidth] {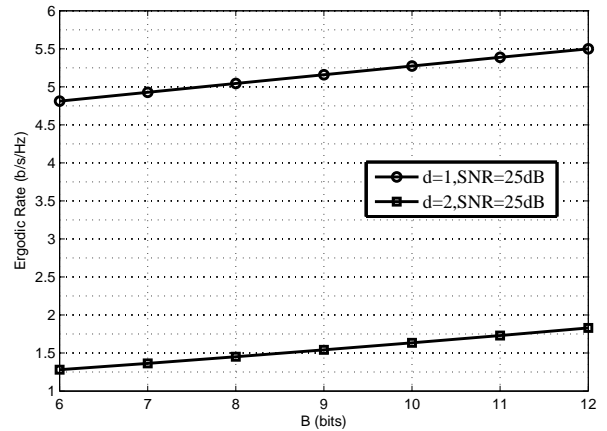}
\caption {Ergodic rate performance at high SNR with large $B$.}
\label{Fig9}
\end{figure}

Next, we show the ergodic rate performance of IA with imperfect CSI.
Similarly, the CSI exchange is useless for performance improvement
at low SNR, as seen in Fig.\ref{Fig4}, which confirms the Theorem 3.
As SNR increases, the performance ceiling appears, if there is
finite CSI exchange amount, resulting an obvious performance gap
with respect to the case of perfect CSI ($B=\infty$), as shown in
Fig.\ref{Fig5}. Note that the number of data stream for each pair
also has a great impact on the ergodic rate. The larger number of
data streams, the more spatial multiplexing gains, but also leads to
higher residual interference. As illustrated in Fig.\ref{Fig6}, at
high SNR, the case of $d=1$ has a significant advantage over that of
$d=2$, since under interference-limited condition, a small spatial
multiplexing is beneficial, which has also been observed in a
multiuser downlink network\cite{ModeSelection}. Although it is
proven that IA can achieve full DOF at high SNR, but it is optimal
to select single data stream from the perspective of maximizing the
ergodic rate under the case of imperfect CSI. Furthermore,
Fig.\ref{Fig9} shows the ergodic rate at high SNR with large $B$. It
is found that the ergodic rate is a linear function of $B$, which
reconfirms the claim of proposition 3.

\begin{figure}[h] \centering
\includegraphics [width=0.5\textwidth] {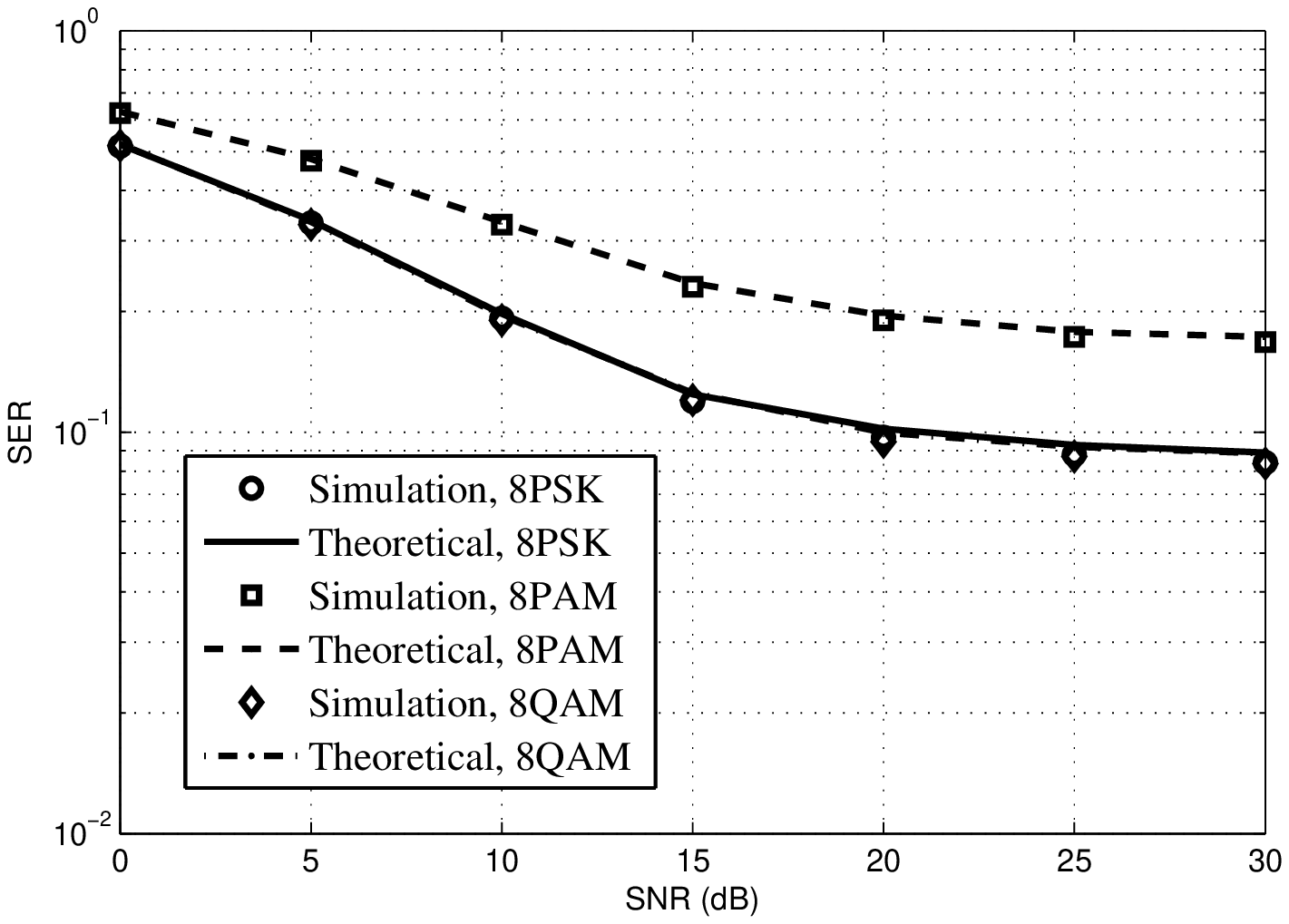}
\caption {Theoretical and simulation SER performance with different
modulation formats and $B=6$.} \label{Fig7}
\end{figure}

\begin{figure}[h] \centering
\includegraphics [width=0.5\textwidth] {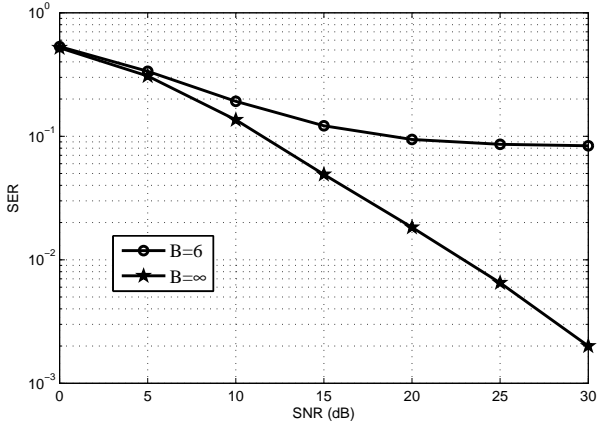}
\caption {SER performance loss due to imperfect CSI with 8QAM
modulation format.} \label{Fig8}
\end{figure}

Finally, we discuss the SER performance of IA with imperfect CSI.
Clearly, for the three different modulation formats, namely 8PSK,
8PAM and 8QAM, the theoretical analysis nicely coincides with the
numerical simulation throughout the whole SNR region, as shown in
Fig.\ref{Fig7}. Imperfect IA caused by imperfect CSI inevitably
leads to error floor, and the performance gap with respect to the
ideal case of perfect CSI becomes larger as SNR increases, as seen
in Fig.\ref{Fig8}. On the other hand, the gap at low SNR is
negligible, which proves that the CSI exchange is useless in such a
scenario.

\section{Conclusion}
This paper comprehensively analyzes the performance of IA with
imperfect CSI over a general MIMO interference channel, including
outage probability, ergodic rate and SER. Thus, the exact
performance can be evaluated for a given transmit SNR, channel
condition, and amount of CSI exchange. Moreover, through asymptotic
analysis on the performance loss due to imperfect CSI, we get some
important design guidelines. For example, CSI exchange is useless at
low SNR, and the amount of CSI exchange should be increased
logarithmically to the SNR and linearly proportional to the number
of antennas at high SNR in order to avoid performance saturation.
Though the number of data streams $d$ should be large from the DoF
point of view, we show that having $d=1$ is optimal at high SNR, as
there exists residual interference under imperfect CSI.

\begin{appendices}
\section{Proof of Theorem 2}
Firstly, we prove there always exists a performance floor in terms
of outage probability for IA with imperfect CSI. As
$\Upsilon\rightarrow\infty$, the corresponding outage probability in
(\ref{eqn12}) is transformed as
\begin{eqnarray}
P_{k,j}^{out,high}&=&1-\sum\limits_{i=1}^{K}\sum\limits_{t=1}^{\eta_{k,i}}\Xi_{K}\Bigg(i,t,\{\eta_{k,q}\}_{q=1}^{K},\left\{\frac{\varrho_{k,q}}{\eta_{k,q}}\right\}_{q=1}^{K},\nonumber\\&&\{l_{k,q}\}_{q=1}^{K-2}\Bigg)
\left(1+\frac{\alpha_{k,q}\rho_{k,q}\gamma_{th}}{\alpha_{k,k}\eta_{k,q}}\right)^{-t}.\label{app1}
\end{eqnarray}
It is found that $P_{k,j}^{out,high}$ is independent of SNR, so
given CSI accuracy and channel condition, there is always a
performance floor.

On the other hand, examining the outage probability with perfect CSI
in (\ref{eqn28}), it is a decreasing function of SNR, so the
performance loss becomes larger as SNR increase. When
$\Upsilon\rightarrow\infty$, $P_{k,j}^{out,perfect}$ is equal to
zero, so the maximum performance loss is given by $\Delta
P_{k,j}^{out,\max}=P_{k,j}^{out,high}$, which proves the Theorem 2.

\section{Proof of Proposition 1}
As seen in (\ref{eqn35}), in order to keep a constant gap, the only
way is to add the CSI exchange amount, so that $\Upsilon\rho_{k,q}$
is constant. Then, as $\Upsilon\rightarrow\infty$, the term
$\left(\frac{\varrho_{k,q}\gamma_{th}}{\kappa_{k,k}\eta_{k,q}}\right)^{-t}$
tends to zero, and hence $\Delta P_{k,j}^{out,high}$ is equal to
zero. In other words, $2^{\frac{B_{k,q}}{N_tN_r-1}}$ should at least
increase linearly proportional to $\Upsilon$, namely
$B_{k,q}\sim(N_tN_r-1)\log_2(\Upsilon)$. So we have
$B_k=\sum\limits_{q=1}^{K}B_{k,q}\sim K(N_tN_r-1)\log_2(\Upsilon)$.

\section{Proof of Theorem 4}
If SNR is high enough, the constant 1 of $\gamma_{k,j}$ in
(\ref{eqn6}) is negligible, so the ergodic rate at high SNR is
transformed as
\begin{eqnarray}
R_{k,j}^{high}&=&\frac{1}{\ln(2)}\bigg(E\left[\ln\left(\kappa_{k,k}\left|\textbf{v}_{k,j}^{H}\textbf{H}_{k,k}\textbf{w}_{k,j}\right|^2+I_{k,j}\right)\right]\nonumber\\
&&-E\left[\ln\left(I_{k,j}\right)\right]\bigg)\nonumber\\
&=&\frac{1}{\ln(2)}\bigg(\sum\limits_{i=1}^{L}\sum\limits_{t=1}^{\omega_{k,i}}\Xi_{L}\bigg(i,t,\{\omega_{k,q}\}_{q=1}^{L},\left\{\frac{\xi_{k,q}}{\omega_{k,q}}\right\}_{q=1}^{L},\nonumber\\&&\{l_{k,q}\}_{q=1}^{L-2}\bigg)\left(\psi(t)+\ln(\omega_{k,q})-\ln(\xi_{k,q})\right)\nonumber\\
&&-\sum\limits_{i=1}^{K}\sum\limits_{t=1}^{\eta_{k,i}}\Xi_{K}\bigg(i,t,\{\eta_{k,q}\}_{q=1}^{K},\left\{\frac{\xi_{k,q}}{\eta_{k,q}}\right\}_{q=1}^{K},\nonumber\\&&\{l_{k,q}\}_{q=1}^{K-2}\bigg)\left(\psi(t)+\ln(\eta_{k,q})-\ln(\xi_{k,q})\right)\bigg),\label{app2}
\end{eqnarray}
where $\xi_{k,i}=\alpha_{k,i}\rho_{k,i}/d_i, \forall i\in[1,K]$,
$\xi_{k,L}=\alpha_{k,k}/d_k$ and $\psi(x)=\frac{d\ln(x)}{dx}$ is the
Euler's function. (\ref{app2}) is obtained based on [36,
Eq.4.352.1]. Clearly, at high SNR, the ergodic rate $R_{k,j}^{high}$
is independent of $\Upsilon$, so there is a performance ceiling.
Additionally, the ergodic rate with perfect CSI at high SNR can be
expressed as
\begin{eqnarray}
R_{k,j}^{perfect,high}&=&\frac{1}{\ln(2)}E\left[\ln\left(1+\Upsilon\alpha_{k,k}\left|\textbf{v}_{k,j}^{H}\textbf{H}_{k,k}\textbf{w}_{k,j}\right|^2\right)\right]\nonumber\\
&\approx&\frac{1}{\ln(2)}E\left[\ln\left(\Upsilon\alpha_{k,k}\left|\textbf{v}_{k,j}^{H}\textbf{H}_{k,k}\textbf{w}_{k,j}\right|^2\right)\right]\label{app3}\\
&=&\frac{\ln(\Upsilon\alpha_{k,k})-C}{\ln(2)},\label{app4}
\end{eqnarray}
where $C=0.57721566490$ is the Euler's constant. (\ref{app3}) drops
the constant 1 and (\ref{app4}) is obtained based on [36,
Eq.4.331.1]. Then the performance loss at high SNR is given by
\begin{eqnarray}
\Delta
R_{k,j}^{high}&=&R_{k,j}^{full,high}-R_{k,j}^{high}\nonumber\\
&=&\log_2(\Upsilon\alpha_{k,k})-C-\frac{1}{\ln(2)}\bigg(\sum\limits_{i=1}^{L}\sum\limits_{t=1}^{\omega_{k,i}}\nonumber\\&&\Xi_{L}\left(i,t,\{\omega_{k,q}\}_{q=1}^{L},\left\{\frac{\xi_{k,q}}{\omega_{k,q}}\right\}_{q=1}^{L},\{l_{k,q}\}_{q=1}^{L-2}\right)\nonumber\\
&&\times\left(\psi(t)+\ln(\xi_{k,q})-\ln(\omega_{k,q})\right)+\sum\limits_{i=1}^{K}\sum\limits_{t=1}^{\eta_{k,i}}\nonumber\\&&\Xi_{K}\left(i,t,\{\eta_{k,q}\}_{q=1}^{K},\left\{\frac{\xi_{k,q}}{\eta_{k,q}}\right\}_{q=1}^{K},\{l_{k,q}\}_{q=1}^{K-2}\right)\nonumber\\
&&\times\left(\psi(t)+\ln(\xi_{k,q})-\ln(\eta_{k,q})\right)\bigg).\label{app5}
\end{eqnarray}
The performance loss enlarges logarithmically with $\Upsilon$, which
proves the Theorem 4.

\section{Proof of Proposition 3}
Let $B$ be the same CSI exchange amount, if both $\Upsilon$ and $B$
are large enough, the ergodic rate can be approximated as
\begin{eqnarray}
R_{k,j}^{high}&\approx&\frac{1}{\ln(2)}\left(E\left[\ln\left(\frac{\kappa_{k,k}\left|\textbf{v}_{k,j}^{H}\textbf{H}_{k,k}\textbf{w}_{k,j}\right|^2}{I_{k,j}}\right)\right]\right)\nonumber\\
&=&\frac{1}{\ln(2)}\Bigg(E\left[\ln\left(\alpha_{k,k}\left|\textbf{v}_{k,j}^{H}\textbf{H}_{k,k}\textbf{w}_{k,j}\right|^2\right)\right]\nonumber\\
&&-E\Bigg[\ln\bigg(\alpha_{k,k}\rho_{k,k}\|\textbf{h}_{k,k}\|^2\sum\limits_{l=1,l\neq
j}^{d_k}\left|\textbf{e}_{k,k}^H\textbf{T}_{j,l}^{(k,k)}\right|^2\nonumber\\
&&+\sum\limits_{i=1,i\neq
k}^{K}\alpha_{k,i}\rho_{k,i}\|\textbf{h}_{k,i}\|^2\sum\limits_{l=1}^{d_i}\left|\textbf{e}_{k,i}^H\textbf{T}_{j,l}^{(k,i)}\right|^2\bigg)\bigg]\Bigg)\nonumber\\
&=&\frac{\ln(\alpha_{k,k})-C}{\ln(2)}-\frac{1}{\ln(2)}\sum\limits_{i=1}^{K}\sum\limits_{t=1}^{\eta_{k,i}}\nonumber\\&&\Xi_{K}\left(i,t,\{\eta_{k,q}\}_{q=1}^{K},\left\{\frac{\xi_{k,q}}{\eta_{k,q}}\right\}_{q=1}^{K},\{l_{k,q}\}_{q=1}^{K-2}\right)\nonumber\\
&&\times\left(\psi(t)+\ln(\xi_{k,q})-\ln(\eta_{k,q})\right)\nonumber\\
&=&\frac{B}{N_tN_r-1}+\frac{\ln(\alpha_{k,k})-C}{\ln(2)}-\frac{1}{\ln(2)}\sum\limits_{i=1}^{K}\sum\limits_{t=1}^{\eta_{k,i}}\nonumber\\&&\Xi_{K}\left(i,t,\{\eta_{k,q}\}_{q=1}^{K},\left\{\frac{\xi_{k,q}}{\eta_{k,q}}\right\}_{q=1}^{K},\{l_{k,q}\}_{q=1}^{K-2}\right)\nonumber\\
&&\times\left(\psi(t)+\ln(\alpha_{k,q})-\ln(\eta_{k,q})\right),\label{app6}
\end{eqnarray}
where (\ref{app6}) holds true because of
$\xi_{k,q}=\alpha_{k,q}\rho_{k,q}/d_q$ and follows the fact of
$\sum\limits_{i=1}^{K}\sum\limits_{t=1}^{\eta_{k,i}}\Xi_{K}\bigg(i,t,\{\eta_{k,q}\}_{q=1}^{K},\\\left\{\frac{\xi_{k,q}}{\eta_{k,q}}\right\}_{q=1}^{K},\{l_{k,q}\}_{q=1}^{K-2}\bigg)=1$.
So the performance ceiling is a linear function of $B$. Given the
channel conditions, the performance gain by adding the CSI exchange
amount from $B_1$ to $B_2$ is equal to $\frac{B_2-B_2}{N_tN_r-1}$,
which is linear proportionally to the increment of CSI.

\end{appendices}

%\begin{biography}[{\includegraphics[width=0.9\textwidth] {Xiaoming_Chen.eps}}]{Xiaoming Chen}
%[M'10-SM'14] received his B.Sc. degree from Hohai University in
%2005, M.Sc. degree from Nanjing University of Science and Technology
%in 2007 and Ph. D. degree from Zhejiang University in 2011, all in
%electronic engineering. He is now with the College of Electronic and
%Information Engineering, Nanjing University of Aeronautics and
%Astronautics, Nanjing, China. His research interests mainly focus on
%cognitive radio, multi-antenna techniques, wireless security,
%interference network and wireless power transfer, etc.
%\end{biography}
%
%\begin{biography}[{\includegraphics[width=0.9\textwidth] {Chau_Yuen.eps}}]{Chau Yuen}
%[SM'12] received the B.Eng. and Ph.D. degrees from Nanyang
%Technological University, Singapore, in 2000 and 2004, respectively.
%In 2005, he was a Postdoctoral Fellow with Lucent Technologies Bell
%Labs, Murray Hill, NJ, USA. In 2008, he was a Visiting Assistant
%Professor with Hong Kong Polytechnic University, Kowloon, Hong Kong.
%From 2006 to 2010, he was with the Institute for Infocomm Research,
%Singapore, as a Senior Research Engineer. Since 2010, he has been an
%Assistant Professor with the Singapore University of Technology and
%Design, Singapore. Dr. Yuen serves as an Associate Editor for the
%IEEE TRANSACTIONS ON VEHICULAR TECHNOLOGY. He received the IEEE
%Asia-Pacific Outstanding Young Researcher Award in 2012.
%\end{biography}

\end{document}